\documentclass[a4paper]{jpconf}
\usepackage{graphicx}
\usepackage{mcite}
\usepackage{xcolor}
\usepackage{listings}
\usepackage{color}
\usepackage{hyperref}

\definecolor{mygreen}{rgb}{0,0.6,0}
\definecolor{mygray}{rgb}{0.5,0.5,0.5}
\definecolor{mymauve}{rgb}{0.58,0,0.82}

\begin{document}
\title{Deployment of High Energy Physics software with a standard method }

\author{Thomas Hahn and Andrii Verbytskyi}

\address{Max-Planck Institute for Physics (Werner Heisenberg Institut), F\"oringer Ring 5, M\"unchen 80805, DE}

\ead{andrii.verbytskyi@mpp.mpg.de}

\begin{abstract}

The installation and maintenance of scientific software for research in
experimental, phenomenological, and theoretical High Energy Physics (HEP)
requires a considerable amount of time and expertise. While many tools are
available to make the task of installation and maintenance much easier,
many of these tools require maintenance on their own, have little
documentation and very few are used outside of HEP community.

For the installation and maintenance of the software, we rely on the well
tested, extensively documented, and reliable stack of software management
tools with the RPM Package Manager (RPM) at its core. The precompiled
HEP software packages can be deployed easily and without detailed Linux
system knowledge and are kept up-to-date through the regular system
update process. The precompiled packages were tested on multiple
installations of openSUSE, RHEL clones, and Fedora. As the RPM
infrastructure is adopted by many Linux distributions, the approach can
be used on more systems.

In this contribution, we discuss our approach to software deployment in
detail, present the software repositories for multiple RPM-based Linux
distributions to a wider public and call for a collaboration for all the
interested parties.

\end{abstract}

\section{Introduction}
The modern studies in theoretical physics, experimental and
 phenomenology of High Energy physics 
(HEP) require a significant amount of specialised software.
The creation of suitable computing environment and setup of required software 
requires significant efforts, time, which poses   a barrier for newcomers 
even if the software is intended to be used unmodified.

In the recent years, these problems were attempted to be solved with a creation of 
different isolated environments, software managing tools or different combinations of those.

However, the available solutions have multiple disadvantages and practically not always
 result in a sufficient reduction of maintenance efforts, require users to learn each solution 
from a scratch and overwhelmingly lack of at least some important features.

Therefore, in our work, we present an approach which 
 avoids implementation of custom software deployment system or isolated environments.
Instead, we concentrate on the practical application and development of the existing ``standard''  solutions of 
the software deployment for the purposes of theoretical, experimental and phenomenology HEP computing.

In our contribution we show the application of this approach in the Linux/GNU OS distributives of
RedHat and SUSE families using corresponding standard tools and present 
a repository with HEP software -- HEPrpms.

\section{Overview of software deployment techniques}
In the recent years, a number of different approaches to the software deployment has been developed.

The most widely used are discussed below with advantages and disadvantages 
of each approach given for the use case of of performing phenomenology studies 
in HEP-phenomenology/HEP-theory.

\subsection{Scripted installation}
The installation with scripts in  isolated environments is, 
perhaps the oldest approach to the software installation.
The installation is typically done with a shell `bootstrap' script on Linux or 
MacOS  with all the dependencies compiled on the spot. In the recent years this approach is often mixed with the 
Python virtual environments~\cite{pyenv}, which enhances the isolation of the 
resulting software from the standard environment of the host system.

The advantage of this approach is a relative simplicity of the setup and 
relatively uniform behaviour across different Unix-like operating systems.
The approach also has a high reliability, as in most cases the `bootstrap' scripts are 
created, documented and supported by the authors of the software. Typically, the installation 
does not require  administrator privileges and multiple installations of the same software can exist.
 
The disadvantages include poor integration with the system installed software,
 the absence of binary packaging, debug information,
 version control, automatic updates, reproducibility of issues and tremendous real time/CPU costs.
\subsection{LCG software stack}
The second widespread approach to the software in the HEP community is the usage of LCG software stack~\cite{Roiser:2010zz}.
In this approach the precompiled software is taken as is from 
the LCG stack installed in {\sc CVMFS}~\cite{cvmfs}. The advantage of this approach is that the LCG software stack
is quite complete, it
is tested by the LHC experiments in collaboration with software developers, regularity obtains
updates and no user maintenance is required beyond the availability of the {\sc CVMFS} file system.
The disadvantage is  a need to maintain 
the complicated scripts that 
required to set-up any non-trivial environment, poor compatibility
with non-LCG software, requirement of the presence of {\sc CVMFS},
and absence of useful debug information. In some cases a specific knowledge of 
the internal details on the software configuration (e.g.\  applied patches) in LCG is needed.
\subsection{``Universal'' package managers}
The third approach is the usage of different ``universal'' package managers.
To this category one can include the installation of software with {\sc PyPi}\cite{pypi}/ 
{\sc Anaconda}\cite{anaconda} and  the installation with 
{\sc spack}~\cite{Volkl:2021yvp}. 

The main advantage of {\sc PyPi}\cite{pypi}/{\sc Anaconda}\cite{anaconda}
 approach is cross-platforming of these package managers and the 
simplicity of usage -- 
 all those package managers 
concentrate on the installation experience of the users, nice documentation 
and the availability of binary packages in public repositories. The installation 
does not require  administrator privileges and multiple installations of the same 
software can co-exist.

 The disadvantages of this approach include  the maintenance of the software stack is 
 not centralised and quite often is semi-anonymous, the stack lacks debug packages
 and requires a maintenance of installations scripts/configurations etc.
 A sizeable disadvantage  is also lack of important 
 software functionality which arises from the lack of dependencies available in the  
 {\sc PyPi}/{\sc Anaconda} repositories
 \footnote{ For instance, the recent  tremendous effort  to provide ROOT~\cite{Antcheva:2011zz} 
 in the {\sc Anaconda} does not include the functionality related to Grid, 
 as the dependencies for this functionality rely on the stack of libraries native 
 for RedHat and Debian systems~\cite{Ellert:2007faf} and an introduction of the stack in the
 anaconda would require a huge effort on itself.}.

The advantage of installation of software with {\sc spack} is the ability to easily 
specify all the dependencies that are used in the installation process, e.g.\  explicitly specify the exact
version of the compiler and the dependency libraries\footnote{
In practice this advantage is very tiny, as many
software packages relevant for HEP don't include support of operating systems 
other than Linux (or less frequently MacOS) and support of compilers other than GNU.
%The support of multiple versions of GNU compiller chains
} and have multiple installations without exercising root privileges.
From the non-technical point of view, an important advantage of 
{\sc spack} is the presence of large community.

The disadvantages include the absence of public binary repositories,
missing debug information, a requirement to 
learn the {\sc spack} itself for using non-trivial software stacks, absence of automatic updates
 and the huge CPU costs for each and every installation from the sources.
\subsection{Containers}
A higher level of abstraction of the deployment approaches
 is the deployment of software in containers.
We do not consider it an independent deployment approach, 
as in the end the installation process 
of software in a container uses one of the approaches already 
described above and has the same advantages/disadvantages.
\subsection{The suggested approach -- HEPrpms}
The software deployment in the  HEP phenomenology/theory/analysis studies outside of large collaborations
   process should have 
some specific properties.
Most important should -- it should have low maintenance costs -- a dedicated effort 
for the development of group-specific tools for deployment is excluded. In the same time 
the approach should be reliable and the deployment should be reproducible. 
While the demands are high, the solution is quite simple.
 The task of software deployment is not new and was successfully solved in
 the software industry in multiple ways.
 
Therefore, we choose the approach of using the software deployment process as it is
recommended by the OS vendors. Practically, this means one should implement the 
approach of the software deployment suggested by the OS vendors most widely used in
HEP.
The overwhelming majority of the computing systems used in HEP are the installations of 
RedHat, SUSE or Debian families of Linux with a small number of Windows and MacOS installations.
Both RedHat and SUSE flavours of Linux have almost identical approaches to the installation
of software which is based on the RedHat Package Manager (RPM)~\cite{rpm}.
Therefore deployment of software with RPMs seems to be quite an attractive approach.
The packaging process is explained in detail in the documentation 
of Fedora or SUSE and is briefly described below.

The packaging requires two  components: the sources of the package
and the {\sc .spec} file. The later is a recipe how to build the package. It has a 
very standardised sections with meta-information on the package version, purpose, used sources, 
required software dependencies, the list of files to be installed.
The {\sc .spec} files (see Listing~\ref{sp} as an example) are split into sections which correspond to the
preparation for build, build, installation and which are executed separately
after the expansion of the instructions in the {\sc .spec} files.
 The instructions in {\sc .spec} files are given
with many convenience macros listed in the Fedora documentation~\cite{frpms}, however many can be given 
as simple shell commands, e.g.\ the  expansion of the sources
can be done just with one command 
setup, see Listing~\ref{sp}.
The {\sc .spec} file and the corresponding sources are given to the standard
{\sc rpmbuild} command which is used to build a source RPM package {\sc .src.rpm} or
 to build the source package and the the binary package. 
The resulting packages can be either directly installed to the desired system 
or put into a repository and the installed from the repository using {\sc yum} or {\sc dnf} utilities.
\begin{lstlisting}[language=bash,label=sp,caption={A spec file for the TheP8I package.}]
Name:           TheP8I
Version:        2.0.1
Release:        3%{?dist}
License:        GPL
Url:            https://gitlab.cern.ch/TheP8I/TheP8I
Source0:        https://gitlab.cern.ch/TheP8I/TheP8I/-/archive/2.0.1/%{name}-2.0.1.tar.gz
Prefix:         %{_prefix}
Summary:        Lund hadronisation for Herwig
%if 0%{?rhel} || 0%{?fedora}
BuildRequires:  pythia8-devel pythia8 gsl lhapdf lhapdf-devel
Requires:       pythia8 gsl ThePEG
%endif
%if 0%{?suse_version}
BuildRequires:  pythia-devel libpythia8 gsl libLHAPDF LHAPDF-devel
Requires:       libpythia8 gsl ThePEG
%endif
BuildRequires: gcc-c++ gcc ThePEG-devel ThePEG autoconf automake libtool gsl-devel
%description
Lund hadronisation for Herwig. Part of earlier ThePEG codes.

%prep
%setup -qn TheP8I-2.0.1

%build
%configure --disable-rpath 
make %{?_smp_mflags}
%install
%make_install

%files
%{_libdir}/ThePEG/*
%_datadir/*

%changelog
* Sun Aug 01 2021 Andrii Verbytskyi<andrii.verbytskyi@mpp.mpg.de> 2.0.1
  - RPATH
* Thu Mar 12 2020 Andrii Verbytskyi <andrii.verbytskyi@mpp.mpg.de> 2.0.0
+ Update to 2.0.0
\end{lstlisting}

{\sc rpmbuild}, {\sc yum} and {\sc dnf}  are standard tools, with documentation on the build process 
provided by the  RedHat and SUSE, therefore there is  no burden of building tools maintenance and
 the preparation to the software packaging amounts to
the creation of the {\sc .spec} files. 

The final product, HEPrpms software repository, is implemented as a set of 
{\sc .spec} files prepared according to the standards of 
Fedora and SUSE. Those {\sc .spec} files are available in a {\sc git} 
repository \url{https://github.com/andriish/HEPrpms} to which any interested person can contribute.
In addition, the git repository includes a trivial script that can grab the sources from the locations
set in the {\sc .spec} files, checks the {\sc md5sums} of the software sources and 
combines the sources with {\sc .spec} files into {\sc .src.rpm} packages.
The software building can be done then from the {\sc .src.rpm} files manually, however, a more attractive approach is to 
create the software packages in specialised services as {\sc CBS}, {\sc copr} etc.

As of 2022 the set of the {\sc .spec} files is used in the {\sc copr} build service to create a set of
rpms.  The corresponding repository is available at\\ \url{https://copr.fedorainfracloud.org/coprs/averbyts/HEPrpms}.
The repository requires quite little manpower to maintain, ${\cal O}(1h/month)$, however 
the updates of the repository are not regular and dedicated to updates of the software in the repository.

\section{Advantages and disadvantages}

The first obvious disadvantage of the  suggested approach is the 
compatibility only with the RedHat and SUSE Linux systems.
This stands in a contrast with the ``universal'' package managers approach which can
provide support for multiple operating systems.
However, practically many software packages are supported only for Linux and MacOS anyway.

Another perceived disadvantage could be the problems with building  certain software with  a 
specialised, i.e\ non-system compiler.
The later capability is provided by e.g.\ {\sc spack}.
While using a different compiler chain could be easily implemented in the HEPrpms approach,
 we argue that often only  GNU compilers are well supported.
And while using a much  newer {\sc gcc} version could have some benefits, those benefits 
are negligible in case of modern operating systems.
The other, related disadvantages -- the requirement of root privileges for the installation 
and ability to use only one particular version of software are discussed in the next section.

The advantages of the approach are that the package system is already maintained by the 
distributives for more than 20 years and is extremely reliable, stable and very well tested. 
The documentation is included in each and every release of RedHat/SUSE Linux distribution.
It does not require extra maintenance manpower as the {\sc spack}, {\sc nixpkgs}, etc.\  for the build infrastructure.

The second advantage is the availability of the binary repositories for the packages 
and the debug information.
The binary packages can be provided by the {\sc pip} package manager and with some extra efforts by the {\sc spack/nixpkg},
however none of them provide the repositories with the debug information.
Without the available binary repositories the usage of those installation methods for the 
CI is much more difficult in comparison to HEPrpms. The availability of the 
debug packages compatible with the system installed software makes the HEPrpms 
the only reliable option for the debug of applications.

The third most important advantage is that from the point of view of user, there is exactly zero
maintenance time, knowledge requirements for the installation of software from HEPrpms.
In this respect only the LCG stack stands on the same level as HEPrpms.

The fourth major advantage is the availability of cloud services that support the RPM approach:
the very long standing OBS, CBS and COPR.
\section{HEPrpms and containers }
The two disadvantages of the RPMs based approach named in the previous section -- the 
requirements of root privileges and the inability to have multiple versions of the same software 
are completely fixed with the capabilities provided by
modern container engines. Consequently, standard packaging 
 is the most practical approach to software deployment/maintenance for the continuous integration (CI) workloads.
 Using a standard container with precompiled packages from the HEPrpms repository does not require 
 the creation of custom container images and therefore 
 brings the costs of the CI containers creation and updates to zero.
E.g.\ even the installation of software in a generic container from the vendors
is fast enough for most of the CI workflows\footnote{This example
could be extended with a simple caching mechanism to reduce the network I/O.
}, see Listing~\ref{fcont}.
\begin{lstlisting}[language=bash,caption=Example of GitLab CI (in YAML) configuration with a generic container from Fedora.,label=fcont]
stages:
  - test
Fedora35-gcc-generic:
  image: fedora:35
  before_script:
    - uname -a 
    - dnf -y install dnf-plugins-core
    - dnf -y copr enable averbyts/HEPrpms 
    - dnf -y install  TheP8I TheP8I-devel
  script:
#Here one can put all the commands that use TheP8I
    - ls
  stage: test
\end{lstlisting}

\section{Conclusions}
The idea to use the standard approach for the HEP-related software deployment
is not new per se and was used many times in the past for the individual 
packages as well as for the larger projects~\cite{Ellert:2007faf}. However, the novelty 
of the presented work is that it shows an example how a reasonable large set of 
software can be packaged and prepared for a very easy deployment using 
the very well developed tools from the modern Linux/GNU OS and the available cloud services.

\section{Acknowledgments}
%Authors wishing to acknowledge assistance or encouragement from 
Thanks to Matthias Ellert for a great example of HEP software packaging.

\section*{References}
%\bibstyle{iopart-num}{JPCSLaTeXGuidelines}
{\bibliographystyle{./iopart-num}{\raggedright\bibliography{HR.bib}}}\vfill\eject
%\begin{thebibliography}{9}
%\bibitem{iopartnum} IOP Publishing is to grateful Mark A Caprio, Center for Theoretical Physics, Yale University, for permission to include the {\tt iopart-num} \BibTeX package (version 2.0, December 21, 2006) with  this documentation. Updates and new releases of {\tt iopart-num} can be found on \verb"www.ctan.org" (CTAN). 
%\end{thebibliography}
\end{document}